\documentstyle[prl,aps,epsf,amssymb]{revtex}
\begin{document}
\draft

\twocolumn[\hsize\textwidth\columnwidth\hsize\csname@twocolumnfalse\endcsname

\title{Driven vortices in 3D layered superconductors:
Dynamical ordering along the c-axis}

\author{Alejandro B. Kolton,$^1$ Daniel Dom\'{\i}nguez,$^1$ 
Cynthia J. Olson$^2$ and Niels Gr{\o}nbech-Jensen$^{3,4}$ }
\address{$^1$ Centro At\'{o}mico Bariloche, 8400 S. C. de Bariloche,
Rio Negro, Argentina\\
$^2$ Department of Physics, University of California, Davis, 
California 95616\\
$^3$ Department of Applied Science, University of California,
Davis, California 95616\\
$^4$ NERSC, Lawrence Berkeley National Laboratory,
Berkeley, California 94720}

\date{\today}
\maketitle
\begin{abstract}
We study a 3D model of driven vortices in
weakly coupled layered superconductors with strong
pinning. 
Above the critical force $F_c$, we find  a plastic flow
regime  
in which pancakes  in different layers are uncoupled,
corresponding to a {\it pancake gas}.
At a higher $F$, there is an ``smectic flow'' regime with
short-range interlayer order,
corresponding to an entangled {\it line liquid}. 
Later, the transverse displacements
freeze and vortices become correlated along the c-axis, 
resulting in a  {\it transverse solid}. Finally, at a  
force $F_s$ the
longitudinal displacements freeze and we find a {\it coherent
solid} of rigid lines.

\end{abstract}


]                

\narrowtext

It is well-known that an external current can induce
an ordering of the vortex structure in superconductors with pinning \cite{thorel}.
For a long time, it was believed that the high-current
phase would have crystalline order.
Recently, it has been found that different 
kinds of order are possible at  high currents, depending
on pinning strength and dimensionality 
\cite{KV,theo,exp,plastico,2d,kolton}. 
This has  led
to numerous theoretical \cite{KV,theo}, experimental \cite{exp} and 
numerical studies \cite{plastico,2d,kolton}.
A crystal-like structure, which could be either a 
perfect crystal \cite{KV} or a Bragg glass 
\cite{theo}, is only  possible in  $d=3$ at large drives. 
In $d=2$, or in $d=3$ for intermediate currents,
a transverse glass is expected, with order only
in the direction perpendicular to the driving force \cite{theo,2d,kolton}.
In the equilibrium vortex phase diagram, the behavior  of
vortex line correlations along the direction of the magnetic field
($c$-axis) has been intensively discussed both experimentally \cite{trafo}
and theoretically \cite{ejec}. In the case of 
driven vortices, little is known on how the 
$c$-axis line correlations would behave in the different dynamical 
regimes. Here we will address this issue starting from
the less favorable case: weakly coupled superconducting planes
with strong pinning. We will show how
the order along the $c$-axis and the in-plane structural order  
take place in a sequence of dynamical phases upon increasing current. 

We study pancake vortices in a layered superconductor, considering
the long-range magnetic interactions between all the 
pancakes and neglecting Josephson coupling \cite{clem}. This model 
is adequate when the interlayer periodicity $d$ is much smaller
than the in-plane penetration length $\lambda_\parallel$ \cite{clem}.
Previous simulations of driven vortices in 3D superconductors have been 
performed  using Langevin dynamics of
short-range interacting particles \cite{srld} or the 
driven isotropic 3D XY model \cite{3dxy}.

The equation of
motion for a pancake located in position ${\bf R}_{i}=({\bf
r}_{i},z_i)=(x_{i},y_{i},n_i d)$,
(${\bf z}\equiv{\hat c}$), is:
\begin{equation}
\eta \frac{ d{\bf r_i}}{dt} = \sum_{j\not= i}{\bf F_v}(\rho_{ij},z_{ij})
+\sum_p{\bf F_p}(\rho_{ip}) + \bf{F}\; ,
\end{equation}
where $\rho_{ij}=|{\bf r}_i-{\bf r}_j|$ and $z_{ij}=|z_i-z_j|$ are     
the in-plane and inter-plane distance between pancakes $i,j$, $\rho_{ip}=
|{\bf r}_i-{\bf r}_p|$ is the in-plane distance between the vortex $i$ and
a pinning
site at ${\bf R_p}=({\bf r}_p,z_i)$,
 $\eta$ is the Bardeen-Stephen friction,
 and ${\bf F}=\frac{\Phi_0}{c}{\bf J}\times{\bf z}$
is the driving force due to an in-plane current ${\bf J}$.
We consider a random uniform distribution of attractive pinning centers in
each layer with
${\bf F_p}=-2 A_p e^{-(\rho/a_p)^2} {\bf r}/a_p^2$, where $a_p$
is the pinning range.
The magnetic interaction between pancakes 
${\bf F}_v(\rho,z)=F_{\rho}(\rho,z) \hat{r}$ is
given by \cite{clem,houston}:
\begin{eqnarray}
F_{\rho}(\rho,0)&=&\frac{A_v}{\rho}
\left[1-\frac{\lambda_{\parallel}}{\Lambda}
\left(1-e^{-\rho / \lambda_{\parallel}}\right)\right]\\
 F_{\rho}(\rho,z_n)&=&-\frac{\lambda_{\parallel}}{\Lambda}\frac{A_v}{\rho}
\left[e^{-|z_n|/\lambda_{\parallel}}-e^{-R_n/\lambda_{\parallel}}\right] \; .
\end{eqnarray}
Here, $R=\sqrt{z^2+\rho^2}$
and $\Lambda=2 \lambda_{\parallel}^2/d$ is the 2D thin-film
screening length. An analogous model to Eqs.\ (2-3) was used in \cite{reefman}.
We normalize length scales by $\lambda_{\parallel}$, energy scales by
$A_v=\phi_0^2 / 4 \pi^2 \Lambda$,
and time is normalized by
$\tau=\eta \lambda_{\parallel}^2/A_v$. 
We consider $N_v$ pancake vortices and $N_p$ pinning
centers per layer in $N_l$ rectangular layers of size $L_x\times L_y$,
and the normalized vortex density is 
$n_v=B\ \lambda_{\parallel}^2 /\Phi_0=(a_0/\lambda_{\parallel})^2$.
We consider 
$n_v=0.29$ with $L_y=16\lambda_\parallel$ and 
$L_x=\sqrt{3}/2L_y$, $N_l=8$ and $N_v=64$.
We take a pinning range of $a_p=0.2$, 
a large pinning strengh of $A_p/A_v=0.2$,
with a high density of pinning centers $n_p=3.125 n_v$. 
The model of Eq.(2-3) is valid in the limit 
$d\ll\lambda_\parallel\ll\Lambda$.
We take $d/\lambda_\parallel=0.01$, 
which corresponds to BSCCO compounds \cite{clem}.
Moving pancake vortices induce a total electric field  ${\bf
E}=\frac{B}{c}{\bf v}\times{\bf z}$, with ${\bf v}=\frac{1}{N_v N_l}\sum_i
{\bf v}_i$.
We study the dynamical regimes in the velocity-force         
curve at $T=0$, solving Eq.\ (1) for increasing values of ${\bf F}=F{\bf
y}$ \cite{houston}.
We use periodic boundary
conditions both in the planes and in the 
$z$ direction and interactions between all pancakes in all layers
are considered \cite{houston}. 
The periodic long-range in-plane and
inter-plane interaction is  evaluated using Ref.~\cite{log}.
The equations are integrated with a time step of $\Delta t=0.01\tau$ and
averages are
evaluated in $16384$ integration steps after $2000$ iterations for
equilibration.
Each simulation is
\begin{figure}
\centerline{\epsfxsize=8.5cm \epsfbox{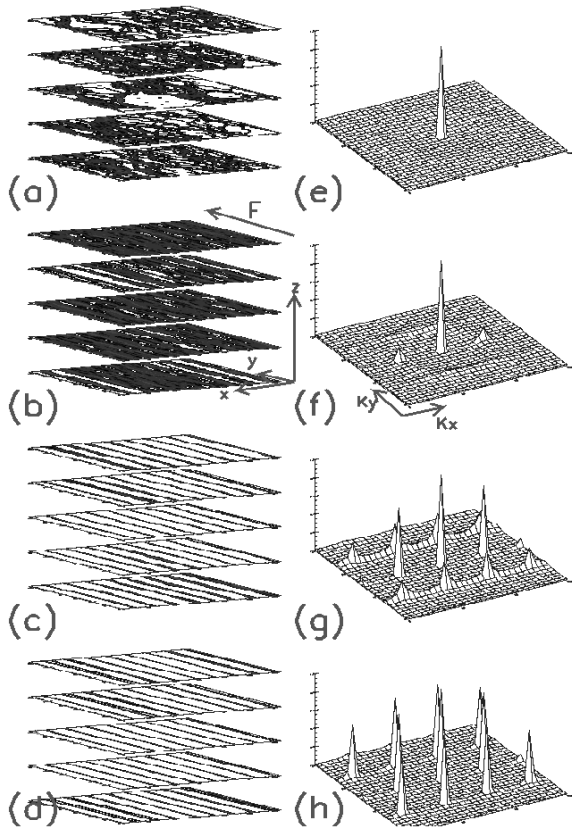}}
\caption{
 Vortex trajectories in the first five layers: (a) F=0.6, (b) F=1.1, (c)
F=2.0. (d) F=3.9. Surface intensity plot of the averaged in-plane 
structure factor S({\bf k}): (e) F=0.6, (f) F=1.1, (g) F=2.0. (h) F=3.9.}
\end{figure}
\noindent
started at $F=0$ with a
triangular  vortex lattice and slowly
increasing the force in steps of $\Delta F= 0.1$ 
up to values as high as $F=8$.

We start with a qualitative description of the different
steady states that arise as a function of increasing force.
In Figure 1(a-d) we show the
vortex
trajectories $\{ {\bf R}_i(t)\}$ for  typical values of $F$ by
plotting the positions of the pancakes in five of the layers for all $t$.
In Fig.1(e-h) we show the average in-plane
structure factor $S({\bf k})= 
\langle\frac{1}{N_l}\sum_n|\frac{1}{N_v}\sum_{i} \exp[i{\bf k}\cdot{\bf
r}_{ni}(t)]|^2\rangle$, with ${\bf k}=(k_x,k_y)$.
Above the depinning critical force $F_c$, we find the following dynamical
phases.
(i){\it Plastic flow} ($F_c<F<F_p$): Pancakes flow in an intrincate
 network of ``plastic'' channels similar to the behavior found in
 2D \cite{plastico,kolton}. 
 The motion in different planes is completely uncorrelated,
 [Fig.1(a)] and there is no signature of order in the structure factor
 [Fig.1(e)].
(ii){\it Smectic flow} ($F_p<F<F_t$): The motion organizes in ``elastic''
channels that are almost parallel and separated by a distance $\sim a_0$,
see Fig.1(b). Small and broad ``smectic'' peaks appear in $S({\bf k})$
for ${\bf k}\cdot{\bf F}=0$ [Fig.1(f)].
There are ``activated'' jumps  of pancakes between channels.  
Along the $c$-direction the channels 
tend to align sitting on top of 
\begin{figure}
\centerline{\epsfxsize=8.5cm \epsfbox{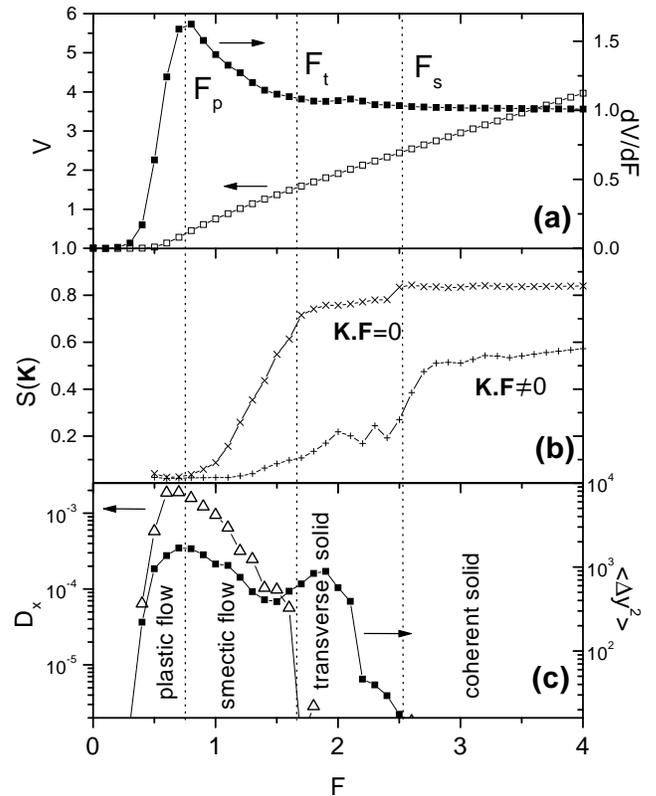}}

\caption{(a) Velocity-force curve, 
left scale, black
points, $dV/dF$ (differential resistance), right scale, white points. (b)
Intensity of the Bragg peaks. For smectic ordering $S(G_1)$, $K_y=0$, $(\times)$
symbols. For longitudinal ordering $S(G_{2,3})$, $K_y \neq 0$, $(+)$ symbols.
(c) Diffusion coefficient for transverse motion $D_x$, $(\bigtriangleup)$,
left scale.  Longitudinal displacements
$\langle[\Delta y(t)]^2\rangle$ for a given $t$ as a function of $F$,
$(\blacksquare)$, right scale.
}  
\end{figure}
\noindent
each other between neighboring planes.
(iii){\it Transverse solid} ($F_t<F<F_s$): There are well defined
channels in all the planes and the pancakes do not jump between channels
[Fig.1(c)]. The structure factor
has sharp smectic peaks and small ``longitudinal'' peaks 
(${\bf k}\cdot{\bf F}\not=0$) have appeared [Fig.1(g)]. 
The location of channels
is correlated in the $c$-axis.
(iv){\it Coherent solid} ($F>F_s$): The channels become more straigth with small
transverse wandering [Fig.1(d)]. The $S({\bf k})$ 
shows well defined peaks for all ${\bf k}$
in the reciprocal lattice [Fig.1(h)].
The dynamical phases (i)-(iii) are similar to the ones found
previously in 2D thin films \cite{kolton}.

The characteristic forces $F_c,F_p,F_t,F_s$ separating the different
dynamical phases are obtained from the analysis of
the in-plane and out of plane structural and dynamical correlations.
We show in Fig.2 the in-plane structure factor 
and temporal fluctuations, 
obtained in the same way as for 2D \cite{kolton}.
In Fig.2(a) we plot the average velocity 
$V=\langle V_y(t)\rangle=\langle\frac{1}{N_vN_l}\sum_{n,i}
\frac{dy_{ni}}{dt}\rangle$,
in the direction of the force as a function of $F$ and its corresponding
derivative $dV/dF$ (differential resistance).  
The force $F_p$ corresponds to the peak in the
differential resistance. We also see a small second maximum in $dV/dF$ for 
a force between $F_t$ and $F_s$ \cite{olson3d}.
In Fig.2(b) we  plot the
magnitude of the peaks in the in-plane structure factor.
We show the peak height at $G_1=2\pi/a_0{\hat x}$, corresponding to
smectic ordering, and the average of the peaks corresponding
to longitudinal ordering 
at ${\bf G}_2=\pm2\pi/a_0(1/2,\sqrt{3}/2)$ and
${\bf G}_3=\pm2\pi/a_0(-1/2,\sqrt{3}/2)$.
We see that at $F_p$ the smectic peak rises up from zero, then at
$F_t$ it reaches an almost constant value and later at $F_s$ it has 
a small jump. The longitudinal peak has a small finite value for
forces above $F_p$, and only at $F_s$  shows a significant increment
towards a large value.
Comparing with the previous 2D results \cite{kolton}, 
we can make the reasonable  assumption 
that for $F_p<F<F_t$ there is only short-range smectic order 
(since there is activated transverse diffusion beteween elastic
channels, see below), for $F_t<F<F_s$ there is probably quasi-long
range smectic order but short range longitudinal order, and
above $F_s$ there is both transversal and longitudinal order 
(quasilong-range or long-range). What is new, compared with the 2D thin
film case \cite{kolton}, is that  above  a force $F_s$ there is a significant
amount of longitudinal order. This may correspond  either to a
moving crystalline phase (if there is long-range order) or to
a moving Bragg glass (if there is quasilong-range order)
\cite{theo}. 
We have verified that, for a given $F>F_s$, 
there is both longitudinal and transversal
order for system sizes of 
$N_l\times N_v=5\times36,5\times64,8\times64,8\times100,10\times100$.
However, a detailed finite size analysis is not possible with these
few small samples.
We complement our discussion of the in-plane physics with the
study of the temporal fluctuations, which are shown in Fig.2(c).
We calculate the
transverse diffusion coefficient $D_x$ from the
average  quadratic  transverse displacements of vortices 
from their center of mass  position $(\bar X_{n},\bar Y_{n}) $, 
$\frac{1}{N_vN_l}\sum_i[x_i(t)-\bar X_{n_i}(t)-x_i(0)+\bar X_{n_i}(0)]^2
\approx D_xt$.   
We find that $D_x$ is maximum at $F_p$ in coincidence with the
peak in $dV/dF$. Below $F_p$ diffusion is through the intrincate
network of plastic channels, above $F_p$ diffusion is through activated
jumps between elastic channels. $D_x$ sharply drops to zero at
$F_t$, indicating that transverse displacements are localized in
the {\it transverse solid} phase \cite{kolton}.
The drift from the center of mass of longitudinal
displacements $\langle[\Delta y(t)]^2\rangle=
\langle[y_i(t)-\bar Y_{n_i}(t)-y_i(0)+\bar Y_{n_i}(0)]^2\rangle$  
is  superdiffusive for $F<F_s$, similar to the results
observed in 2D films \cite{kolton}. 
For $F>F_s$ the longitudinal displacements become frozen in a constant
value $\langle[\Delta y(t)]^2\rangle < a_0/N_l$, as it
is shown in Fig.2(c). 
Since in-plane displacements are localized and there are large transversal and
longitudinal Bragg peaks, we call this phase a {\it coherent solid}.

Let us now discuss how the ordering along the $c$-axis takes place.
We analyze the pair distribution function:
$g(\rho,n)=\frac{L_xL_y}{N_v}\langle \sum_{i\not=j}
\delta(\rho-\rho_{ij}) \delta_{n,n_{ij}} \rangle$.
From $g(\rho,n)$ we define a correlation function along c-axis
$C_z(n)=\lim_{\rho \rightarrow 0} g(\rho,n)$. Short-range ordering
will be given by a 
\begin{figure}
\centerline{\epsfxsize=8.5cm \epsfbox{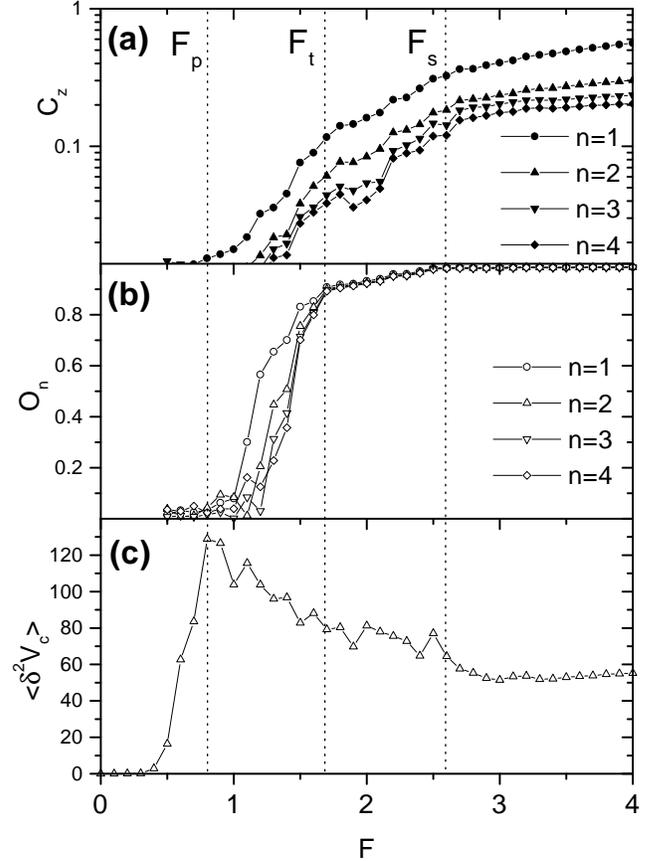}}
\caption{(a) Correlation parameter  
in c-direction of instantaneous configurations $C_z(n)$ vs $F$ for
$n=1,2,3,4$ interplane distance. 
(b) Trajectories overlap correlation parameter in c-direction  
$O_n$ vs $F$ for $n=1,2,3,4$ interplane
distance. 
(c) Voltage fluctuations in c-direction $\langle \delta^2
V_c \rangle$ vs $F$.}   
\end{figure}
\noindent finite $C_z(n=1)$, meaning that pancakes in
neighboring 
planes are coupled and  a ``vortex line'' can therefore
be defined. In principle, an exponential decay $C_z(n)\sim\exp(-n/\xi_z)$
would define a correlation length for the vortex line \cite{ejec}. 
On the other hand, long-range ordering will be given by
$C_z(n\rightarrow\infty)\rightarrow C^{\infty}_z >0$.
In Fig.3(a) we show $C_z(n)$ 
as a function of $F$ for $n=1,2,3,4$. We see that 
at $F_p$ there is an onset of short-range order along
the c-axis with a finite  $C_z(n=1)$.  
At higher forces between $F_p$ and $F_t$ the other $C_z(n>1)$ start to rise. 
The absence of correlations for $F<F_p$ means that
pancake motion is completely random and uncorrelated between different
planes. Therefore, we propose that the plastic flow regime corresponds
to a {\it pancake gas}.
Above $F_p$, in the smectic flow regime, 
it is possible to define a vortex line with short range correlations
along the c-axis. Since there are in-plane jumps between elastic channels
(i.e., cutting and reconnection of flux lines)  
we may consider this  phase as an entangled {\it line liquid}.
Above $F_t$, $C_z(n)$ is finite for all $n$ considered
and  tends to saturate upon increasing $n$. 
This indicates
that vortex lines  become more stiff above $F_t$.
We also analyzed 
the $c$-axis correlation between averaged vortex densities.
We first define 
$\rho_v({\bf r},n,t)=\frac{1}{N_v}
\sum_i \delta({\bf r}-{\bf r}_{ni}(t))$  taking
a coarse-graining scale $\Delta r=a_0/2$  
(results do not vary much for $\Delta r=a_0/4$).
The regions where the average density $\langle\rho_v({\bf r},n)\rangle$ 
is large define the
paths of steady state vortex motion. We can thereby calculate the overlap
function of vortex trajectories between different planes as
$O_n=C_\rho(n)/C_\rho(0)$, with $C_\rho(n)=\frac{L_x L_y}{N_l}
[\sum_m \int d{\bf r} \langle \rho_v({\bf r},m)\rangle \langle \rho_v({\bf
r},m+n) \rangle] - 1$.
This is shown in Fig.3(b). 
We see that $O_n$ also has an onset at $F_p$. For $F_p<F<F_t$,
we have some overlap of the elastic channels that decreases with
increasing $n$, consistent with the entangled line-liquid picture. 
More interestingly, at $F_t$ 
the overlap function $O_n$ becomes independent of $n$. This means that
there is long-range $c$-axis coupling 
of the path of the elastic channels. When transverse displacements 
become localized in the $x$-direction,
they also become locked in the $c$-direction. 
Thus, the freezing of in-plane transverse displacements
occurs simultaneously with a transverse disentanglement of flux lines
at $F_t$.
A striking result is that we find $O_n\approx1$ above $F_s$, i.e., a perfect
c-axis coupling of elastic channels (within the scale $\sim a_0/4$).
Another interesting point to consider is the correlation of
vortex velocities.
If vortices in different planes move at different velocities, they
will induce a Josephson voltage difference along the $c$-axis 
given by $V_{n,n+1}({\bf r},t)=
\frac{\Phi_0}{2\pi c}\frac{d}{dt}\phi_{n,n+1}({\bf r},t)$,
with $\phi_{n,n+1}$ the superconducting phase difference between planes
$n$ and $n+1$. A good approximation  for
pancakes at ${\bf r}_{n,i}$ is to write
$\phi_{n,n+1}({\bf r},t)=\sum_{i}[f({\bf r}-{\bf r}_{n,i})-f({\bf r}-{\bf r}_{n+1,i})]$ 
with $f({\bf r})\approx \arctan(x/y)$.
We can therefore estimate the c-axis voltage fluctuations as 
$\langle \delta^2 V_c\rangle = 
\sum_{n} \int [\langle V^2_{n,n+1}({\bf r},t) \rangle-\langle V_{n,n+1}
({\bf r},t) \rangle^2 ] d{\bf r}\approx
A \sum_{n} [\langle{\bf V}_n^2\rangle -\langle{\bf V}_n\rangle^2]
-[\langle{\bf V}_n\cdot{\bf V}_{n+1}\rangle- 
\langle{\bf V}_n\rangle\cdot\langle{\bf V}_{n+1}\rangle]$; 
with ${\bf V}_n(t) = \frac{1}{N_v}\sum_{i}{\bf v}_{n,i}(t)$,
and the constant $A\sim \log \Lambda$ if $L>\Lambda$ or $A\sim\log(L)$ otherwise.
It is clear that $\langle \delta^2 V_c\rangle=0$ 
for pancakes moving with the same velocity in all planes. 
We see in Fig.3(c) that the voltage
fluctuations have a maximum at $F_p$. 
For $F>F_p$, $\langle \delta^2 V_c\rangle$ 
decreases, and above $F_s$ it reaches an almost $F$-independent value.
The fact that $\langle \delta^2 V_c\rangle$ does not vanish above $F_s$  
is consistent with the result that $C_z(n)<1$ for all values of $F$ in
Fig.3(a). In other words, while transverse displacements are strongly
correlated along the $c$ direction for large forces [Fig.3(b)], 
the longitudinal
displacements in different planes are weakly correlated .

In conclusion, we have clearly distinguished different dynamical phases
in 3D layered superconductors considering both in-plane and $c$-axis 
ordering \cite{olson3d}. 
The onset of short-range $c$-axis correlations could be studied experimentally
with plasma resonance measurements \cite{plasma}. The long-range ordering along
the $c$-axis could be studied through  simultaneous measurements of 
$\rho_c$ resistivity and in-plane
current-voltage response \cite{lamenghi}.

We acknowledge discussions with L.N.\ Bulaevskii, P.S.\ Cornaglia, F.\ de
la Cruz, Y.\ Fasano,
M.\ Menghini.
This work has been supported by  ANPCYT 
(Proy. 03-00000-01034), 
by Fundaci\'on Antorchas
(Proy. A-13532/1-96), Conicet, CNEA and FOMEC
(Argentina);   by CLC and CULAR (Los Alamos), and
by the Director, Office of Adv. Sci. Comp. Res., 
Division of Mathematical, Information, and
Computational Sciences of the U.S.D.O.E. (contract
number DE-AC03-76SF00098).

\end{document}